\documentclass[iop]{emulateapj}
\usepackage{epsfig,amsmath}
\usepackage[dvips]{color}
\usepackage{verbatim}
\usepackage{graphicx}
\usepackage{txfonts}
\usepackage{natbib}
\bibpunct{(}{)}{;}{a}{}{,}
\usepackage{hyperref}

\begin{document}

\title{The properties of the 2175 \AA\ extinction feature discovered in GRB afterglows}

\author{Tayyaba Zafar,\altaffilmark{1,2}
Darach Watson,\altaffilmark{1}
\'{A}rd\'{i}s El\'{i}asd\'{o}ttir,\altaffilmark{1}
Johan P. U. Fynbo,\altaffilmark{1}
Thomas Kr\"{u}hler,\altaffilmark{1,3}
Patricia Schady,\altaffilmark{3}
Giorgos Leloudas,\altaffilmark{1}
P\'all Jakobsson\altaffilmark{4}
Christina  C. Th\"{o}ne,\altaffilmark{5}
Daniel A. Perley,\altaffilmark{6}
Adam N. Morgan,\altaffilmark{6}
Joshua Bloom,\altaffilmark{6}
Jochen Greiner\altaffilmark{3}}

\altaffiltext{1}{Dark Cosmology Centre, Niels Bohr Institute, University of Copenhagen,
Juliane Maries Vej 30, DK-2100 Copenhagen \O, Denmark; tayyaba, darach, ardis@dark-cosmology.dk}
\altaffiltext{2}{Laboratoire d'Astrophysique de Marseille - LAM, Universit\'e Aix-Marseille \& CNRS, UMR7326, 38 rue F. Joliot-Curie, 13388 Marseille Cedex 13, France.}
\altaffiltext{2}{Max-Planck Institut f\"{u}r Extraterrestrische Physik, Giessenbachstrasse, 85748, Garching, Germany.}
\altaffiltext{3}{Centre for Astrophysics and Cosmology, Science Institute, University of Iceland, Dunhagi 5, IS-107 Reykjav\'ik, Iceland.}
\altaffiltext{4}{INAF-Osservatorio Astronomico di Brera, Via Bianchi 46, I-23807, Merate (Lc), Italy.}
\altaffiltext{5}{Department of Astronomy, University of California, Berkeley, CA 94720-3411, USA.}

\begin{abstract}

The unequivocal, spectroscopic detection of the 2175\,\AA\ bump in
extinction curves outside the Local Group is rare. To date, the properties
of the bump have been examined in only two gamma-ray burst (GRB) afterglows
(GRB\,070802 and GRB\,080607). In this work we analyse in detail the
detections of the 2175\,\AA\ extinction bump in the optical spectra of the
two further GRB afterglows: GRB\,080605 and 080805. We gather all available
optical/near-infrared photometric, spectroscopic and X-ray data to construct
multi-epoch spectral energy distributions (SEDs) for both GRB afterglows. 
We fit the SEDs with the \citet{fm2} model with a single or broken
power-law. We also fit a sample of 38 GRB afterglows, known to prefer a
Small Magellanic Cloud (SMC)-type extinction curve, with the same model. We
find that the SEDs of GRB\,080605 and GRB\,080805 at two epochs are fit well
with a single power-law with a derived extinction of
$A_V=0.52^{+0.13}_{-0.16}$ and $0.50^{+0.13}_{-0.10}$, and
$2.1^{+0.7}_{-0.6}$ and $1.5\pm0.2$ respectively. While the slope of the
extinction curve of GRB\,080805 is not well-constrained, the extinction
curve of GRB\,080605 has an unusual very steep far-UV rise together with the
2175\,\AA\ bump. Such an extinction curve has previously been found in only
a small handful of sightlines in the Milky Way (MW). One possible
explanation of such an extinction curve may be dust arising from two
different regions with two separate grain populations, however we cannot
distinguish the origin of the curve. We finally compare the four 2175\,\AA\
bump sightlines to the larger GRB afterglow sample and to Local Group
sightlines. We find that while the width and central positions of the bumps
are consistent with what is observed in the Local Group, the relative
strength of the detected bump ($A_{\rm bump}$) for GRB afterglows is weaker
for a given $A_V$ than for almost any Local Group sightline. Such dilution
of the bump strength may offer tentative support to a dual dust-population
scenario.


\end{abstract}

\maketitle

%
\section{Introduction\label{introduction}}
Starlight in galaxies is absorbed and scattered by dust grains present in the interstellar medium (ISM). The process is usually quantified by the introduction of an interstellar extinction curve. A characteristic feature in the extinction curves of the MW and Large Magellanic Cloud (LMC) is the 2175\,\AA\ extinction bump, first discovered by \citet{stecher65}. The 2175\,\AA\ bump has been attributed to absorption by graphite grains processed by star formation \citep[e.g.,][]{draine03}. However, the exact origin of the 2175\,\AA\ bump is still unclear although several candidates have been suggested ranging from carbonaceous materials \citep{henard} to iron poor silicates in the form of partially hydrogenated amorphous Mg$_2$SiO$_4$ particles \citep{steel}. It has also been suggested that coating on graphite cores can explain the variation in the bump width, and possible candidates for mantle material are a mixture of neutral polycyclic aromatic hydrocarbons (PAHs) or other forms of non-graphitic carbon \citep{mathis94}. 

The most striking characteristics of the 2175\,\AA\ bumps are the remarkably constant central wavelength, large dispersion of height and width, and variable strength, varying from one line of sight to another \citep[see][]{fm3}. The 2175\,\AA\ bump in the MW extinction curves appear to be the strongest known to date, though there are very few absolute extinction curves known outside the Local Group. The feature becomes gradually weaker in the LMC and SMC. The two broad categories of LMC sightlines are LMC-average, having MW-type extinction curves, and LMC2 supershell showing weaker bumps and a steep rise into the UV \citep{nandy,gordon}. The SMC sightlines exhibit a featureless extinction curve and an even steeper rise into the UV. However a line of sight through the SMC wing exhibits an extinction curve with a prominent 2175\,\AA\ bump \citep{lequeux,gordon}. 
 
 \begin{table}
\begin{minipage}[t]{\columnwidth}
\caption{Photometric observations of the afterglow of GRB\,080605. Magnitudes are given in the AB system, with the host contribution subtracted, and corrected for Galactic extinction of $E(B-V)=0.14$.}      
\label{data080605} 
\centering
\begin{tabular}{c c c c c}   
\hline\hline                        
Mid-time &  Exp. time & Instrument & Filters & Magnitudes \\ 
hr & ks & & & AB mag \\
\hline
0.202 & 0.02 & UVOT & $uvw2$ &	$> 19.86$ \\	
6.158 & 0.54 & UVOT & $uvw2$ &	$> 22.52$ \\	
0.194 & 0.04 & UVOT & $uvm2$ &	$> 20.07$ \\
8.375 & 0.26 & UVOT & $uvm2$ &	$> 22.78$  \\
0.201 & 0.04 & UVOT & $uvw1$ &	$> 20.41$ \\
4.546 & 0.50 & UVOT & $uvw1$ &	$> 21.95$  \\
0.208 & 0.04 & UVOT & $u$ &	$> 20.44$  \\	
5.160 & 0.28 & UVOT & $u$ &	$> 21.20$ \\
0.214 & 0.02 & UVOT & $b$ &	$> 19.23$	 \\
7.769 & 0.58 & UVOT & $b$ &	$> 20.13$	 \\
0.113 & 0.39 & UVOT & $v$ &	$18.49\pm0.10$ \\	
0.323 & 0.39 & UVOT & $v$ &	$19.37\pm0.15$ \\
6.772 & 0.24 & UVOT & $v$ &	$> 19.33$  \\
0.042 & 0.10 & UVOT & $white$ & 	$18.87\pm0.08$	\\
0.196 & 0.10 & UVOT & $white$ &	$20.83\pm0.19$	\\
1.555 &  0.14 &GROND & $g^\prime$ &  $20.28 \pm0.07$ \\
2.785 &  1.50 & GROND & $g^\prime$  &  $20.76 \pm0.05$ \\
1.555 &  0.14 & GROND & $r^\prime$  &  $19.69\pm0.06$ \\
2.785 &  1.50 & GROND & $r^\prime$  & $20.15\pm0.05$ \\
5.512 &  3.23 & GROND & $r^\prime$  & $20.68\pm0.05$ \\
6.504 &  3.23 & GROND & $r^\prime$  & $20.82\pm0.06$ \\
1.555 &  0.14 & GROND & $i^\prime$  &  $19.66 \pm0.05$ \\
2.785 &  1.50 & GROND & $i^\prime$  & $19.66\pm0.05$ \\
1.555 &  0.14 & GROND & $z^\prime$  &  $18.93\pm0.07$ \\
2.785 &  1.50 & GROND & $z^\prime$  & $19.35\pm0.05$ \\
1.555 &  0.24 & GROND & $J$  &  $18.40 \pm0.12$ \\
2.785 &  1.20 & GROND & $J$  & $19.01\pm0.09$ \\
6.021 &  6.93 &PAIRITEL & $J$  & $19.44\pm0.22$ \\
9.528 &  9.22 &PAIRITEL & $J$  & $19.39\pm0.21$ \\
1.555 &  0.24 & GROND & $H$  &  $17.99\pm0.15$ \\
2.785 &  1.20 & GROND & $H$  & $18.56\pm0.09$ \\
6.021 &  6.93 &PAIRITEL & $H$  & $18.91\pm0.18$ \\
9.528 &  9.22 &PAIRITEL & $H$  & $19.03\pm0.19$ \\
1.555 &  0.24 & GROND & $K$  &  $17.89\pm0.23$ \\
2.785 &  1.20 & GROND & $K$  & $18.26\pm0.11$ \\
6.021 & 6.93 & PAIRITEL & $K$  & $18.93\pm0.20$ \\
9.528 & 9.22 &PAIRITEL & $K$  & $19.13\pm0.23$ \\
\hline
\end{tabular}
\end{minipage}
\end{table}

\begin{table}
\begin{minipage}[t]{\columnwidth}
\caption{Optical/NIR photometry of the afterglow of GRB\,080805. Magnitudes are given in the AB system and corrected for Galactic extinction of $E(B-V)=0.043$.}      
\label{data080805} 
\centering
\begin{tabular}{c c c c c}   
\hline\hline                        
Time since trigger & Exp. time & Instrument & Filters & Magnitudes \\ 
hr & s & & & AB mag \\
\hline
3.968 & 1104 &  UVOT & $uvw2$ &  $>22.91$ \\
3.958 & 415 &  UVOT & $uvm2$ &  $>21.83$ \\
2.248 & 1082 &  UVOT & $uvw1$ &  $>22.53$ \\
2.404 & 1081 &  UVOT & $u$ &  $>22.11$ \\
0.283 & 150 & VLT/FORS2 & $B$  & $22.72\pm0.04$ \\
0.719 & 150 & VLT/FORS2 & $B$   & $21.16\pm0.06$ \\
3.102 & 1082 &  UVOT & $b$ &  $>21.28$ \\
0.080 & 66 & GROND & $g^\prime$  & $21.66\pm0.10$ \\
0.243  & 40 & VLT/FORS2 & $V$  & $22.15\pm0.04$ \\
0.679  & 40 & VLT/FORS2 & $V$  & $22.87\pm0.08$ \\
3.331 & 1207 &  UVOT & $v$ &  $>20.45$ \\
2.513 & 1180 &  UVOT & $white$ & $>22.87$ \\
0.080 &  66 &GROND & $r^\prime$  & $20.80\pm0.08$ \\
0.150 & 30 & VLT/FORS2 & $R$  & $20.93\pm0.02$ \\
0.322 & 30 & VLT/FORS2 & $R$  &  $21.52\pm0.03$ \\
0.585 & 30 & VLT/FORS2 & $R$  &  $21.78\pm0.04$ \\
0.784 & 10 & VLT/FORS2 & $R$  &  $22.16\pm0.09$ \\
0.822 & 10 & VLT/FORS2 & $R$  &  $22.04\pm0.08$ \\
2.325 & 120 & VLT/FORS2 & $R$  &  $22.95\pm0.06$ \\
0.080 & 66 & GROND & $i^\prime$  & $20.13\pm0.08$ \\
0.216 & 40 & VLT/FORS2 & $I$  &  $20.77\pm0.02$ \\
0.652 & 40 & VLT/FORS2 & $I$  &  $21.52\pm0.05$ \\
0.080 &  66 &GROND & $z^\prime$  & $19.79\pm0.08$ \\
0.180 & 60 & VLT/FORS2 &  $z$-Gunn & $20.09\pm0.02$ \\
0.616 & 60 & VLT/FORS2 & $z$-Gunn &  $21.00\pm0.06$ \\
\hline
\end{tabular}
\end{minipage}
\end{table}

The net attenuation curves of local starburst galaxies show that their dust lacks the 2175\,\AA\ bump \citep{calzetti}. A significant 2175\,\AA\ bump is observed in the spectra of star forming galaxies at $z$ $\sim$ $2$, indicating an LMC-like extinction curve \citep{noll07}. It has also been detected in the Great Observatories Origins Deep Survey (GOODS)-\emph{Herschel} field galaxies at $z$ $>$ $1$ \citep{buat}. The detection of the 2175\,\AA\ bump has been reported in several individual distant absorbing systems \citep[e.g.,][]{junkkarinen,wang,noterdaeme,zhou,jiang}. The feature has also been detected in the SEDs of GRB afterglows with a large diversity of extinction curve shapes \citep{kruhler08,ardis,prochaska09,perley09,zafar}. The detection of the 2175\,\AA\ bump is also reported for an intervening absorber at $z=1.11$ towards GRB\,060418 \citep{ellison}. 

GRBs provide a unique tool for studying the absolute extinction curves of distant galaxies because of their bright afterglow emission, simple power-law spectra and their occurrence in star-forming regions. In this paper we report in detail the observations and analyses of two extinguished GRB afterglows showing a 2175\,\AA\ bump in their optical spectra: GRB\,080605 and GRB\,080805,which we compare to the two spectroscopically-confirmed 2175\,\AA\ bumps in GRB hosts. Previously optical spectra of there afterglows of GRB\,080605 and GRB \,080805 have been presented in \citet{fynbo} and the SEDs have been discussed briefly in \citet{zafar}. Based on photometry, the detection of the bump has also been confirmed for both afterglows \citep{greiner10}.


The paper is organized as follows:
 In \S2 we describe multi-wavelength observations of the afterglows of GRB\,080605 and GRB\,080805 carried out with different instruments. In \S3 we present our
results from the SED fitting. In \S4 we make a comparison between the bump properties of 42 GRB afterglows and Local Group sightlines. We further discuss the extinction curve of GRB\,080605. In \S5 we provide our conclusions. 
\section{Observations and data reduction}\label{observations} 
\subsection{GRB\,080605}\label{080605}
\emph {Swift} Burst Alert Telescope (BAT; \citealt{barthelmy}) triggered on GRB\,080605 on 2008 June 05 at 23:47:57.86 UT. \emph{Swift} X-ray Telescope (XRT; \citealt{burrows}) slew immediately to GRB location and began observing the X-ray afterglow of GRB\,080605. The Gamma-Ray Optical and Near-Infrared Detector (GROND) observed the field in different optical and NIR bands \citep{greiner10}. The afterglow was also observed with the Peters Automatic Infrared Imaging Telescope (PAIRITEL; \citealt{bloom06}) in the $J$, $H$, and $K$ bands starting from 5 to 11 and then 31 hrs after the burst. The GRB is found close to a bright star, which may contribute to the measured flux. We reduced the PAIRITEL data and used image subtraction techniques \citep{alard} to get reasonable photometry of the afterglow. To increase the signal to noise ratio we stacked the images from 5-7 hrs and 8-11 hrs post-burst. The afterglow is clearly detected up to 11 hrs after the burst whereas it is not detected at 31 hrs after the burst (see Table \ref{data080605}). We also reduced the \emph{Swift} Ultra-violet and Optical Telescope (UVOT; \citealt{roming05}) data of the afterglow. After subtracting the contribution from the nearby object we found that the afterglow is not detected in any of the UVOT bands. The host galaxy of GRB\,080605 is bright, with $r^\prime$ $\sim$ $22.8$\,mag \citep[see][]{kruhler11} and the contribution is subtracted from the photometric data. 

An optical spectrum of the afterglow was secured at the Very Large Telescope (VLT) equipped with the FOcal Reducer and low dispersion Spectrograph 2 (FORS2) using the 300V grism at 3.2 hrs (2 exposures with an exposure time of 1200 s each) after the burst \citep{jakobsson083,fynbo}. Spectra were also secured with the 1200R and 1400V grisms yielding a spectral resolution of $R=2140$ and $2100$ respectively. The 300V spectrum was flux calibrated using the spectrum of a spectrophotometric standard star LTT9239 obtained on the same night. The afterglow spectra were taken under good observing conditions. The spectra show several narrow absorption lines with the highest redshift of the absorber at $z_{\rm abs}=1.6403$, which we adopt as the redshift of the GRB. All photometric and spectroscopic data have been corrected for Galactic extinction using the maps of \citet{schlegel}, $E(B-V)=0.14$\,mag.

Ly$\alpha$ absorption could not be detected for this burst with a ground-based telescope due to the relatively low redshift (see \citealt{fynbo} for the optical spectrum). A large number of metal species (e.g. Si, C, Al, Zn, Fe, Mg, Mn and Cr) are identified at the redshift of the GRB \citep{fynbo}. To obtain a limit on the ionic column densities, we analyzed all grism spectra. The absorption features indicate a two component profile in the 1200R and 1400V spectra. The GRB absorption features are saturated and the resolution is not high enough to estimate reliable limits for the metal column density. Assuming the {Mn}\,{\sc ii} $\lambda$2577 line is located at the weak limit of the curve of growth, this would imply that $N$({Mn}\,{\sc ii}) $>13.4$ cm$^{-2}$. The metallicity of the GRB cannot be obtained from the afterglow spectrum due to the absence of Ly$\alpha$ absorption trough. The metallicity inferred from emission lines from the host is around solar \citep{thomas12}. Moreover the equivalent widths of metal lines are compared with \citet{fynbo} and \citet{christensen11} sample. The equivalent widths of metal lines of GRB\,080605 lie above the average for most elements except Fe and Zn. The lower equivalent width for Fe could be due to dust depletion. 

\begin{figure}
	\centering 
	{\includegraphics[width=\columnwidth,clip=]{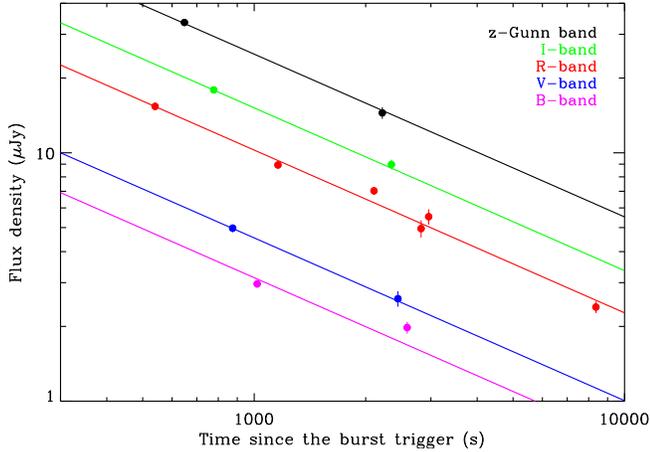} }
	\caption{VLT/FORS2 multi-color lightcurve of the afterglow of GRB\,080805. Solid lines are the decay slope of $\alpha=0.65$, derived from a fit to the $R$-band data. Fluxes are corrected for Galactic extinction of $E(B-V)=0.043$\,mag.} 
		\label{lightcurve} 
\end{figure} 

\subsection{GRB\,080805}\label{080805}
At 07:41:34.73 UT on 2008 August 05 after the BAT trigger, \emph{Swift} XRT began observations of GRB\,080805. GROND observed the afterglow at 4.6 min after the burst in the $g^\prime$, $r^\prime$, $i^\prime$ and $z^\prime$ filters. Further imaging of the afterglow was performed with VLT/FORS2 in the Bessel $B$, $V$, $R$, $I$ and $z$-Gunn filters starting from 9 to 139.5 min after the burst (see Table \ref{data080805}). VLT/FORS2 photometric data were reduced in standard way. The magnitudes of the afterglow were obtained using ESO zero-points from the night of the observation (for $BVRI$ filters) and by observing a standard field covered by the Sloan Digital Sky Survey (SDSS) for the $z$-Gunn band. Fitting the $R$-band afterglow lightcurve provides a temporal index of $\alpha=0.65\pm0.03$ (see Fig. \ref{lightcurve}). 

The optical spectrum of the afterglow was taken with VLT/FORS2 (grism 300V) at 1.0 hr (2 exposures, each of 600 s) after the burst \citep{jakobsson084,fynbo}. The spectrum was flux calibrated using the spectrum of a spectrophotometric standard star LTT1020 observed on the same night. The spectra were also obtained with the 1400V, 1200R and 1028z ($R=2560$) grisms \citep{jakobsson084}. The data were taken under photometric conditions. The UVOT data of the afterglow was reduced and the source was not detected in any of the filters. The redshift is based on a metal system at $z=1.5042$ displaying {Al}\,{\sc ii} $\lambda$1671, {Fe}\,{\sc ii} $\lambda\lambda\lambda$2382, 2586, 2600, {Mg}\,{\sc ii} $\lambda\lambda$2796, 2803 and {Mg}\,{\sc i} $\lambda$2852 absorption lines. All data have been corrected for Galactic extinction using the \citet{schlegel} maps with $E(B-V) = 0.043$\,mag.

The Ly$\alpha$ absorption line is also not seen for this burst due to its relatively low redshift. Because of the low resolution and highly saturated absorption features, ionic column densities could not be obtained for this burst. We compared equivalent widths of metal lines of this burst with \citet{fynbo} and \citet{christensen11} composite samples and find overabundance in Al and Mg. Fe lie about average and this might indicate depletion.

%
%
\begin{table}
\begin{minipage}[t]{\columnwidth}
\caption{Best fit extinction curve parameters of the afterglow SEDs using the FM parameterization. The parameters with fixed values are marked (f).}      
\label{table:1} 
\centering
\begin{tabular}{@{}c c c c c@{}}   
\hline\hline                        
Parameter & \multicolumn{2}{c}{GRB\,080605} & \multicolumn{2}{c}{080805}  \\ 
\hline\hline
 $t_{\rm SED}$ (hr) & 1.55 & 2.78 & 0.08 & 0.72 \\
 $c_1$ &  $-3.33\pm1.59$ & $-6.03\pm0.65$ & $-0.39\pm1.12$ & $0.09\pm0.30$\\
 $c_2$ ($\mu$m)  & $1.92\pm0.55$ &  $2.64\pm0.25$ & $0.63\pm0.38$ &  $0.68\pm0.14$ \\
 $c_3$  & $1.00\pm0.68$ & $0.46\pm0.11$ & $1.39\pm0.63$ & $1.10^{+0.51}_{-0.48}$ \\
 $c_4$ ($\mu$m$^{2}$)  & $1.27$(f) & $1.27\pm0.45$ & $0.40$(f) & $0.40\pm0.16$\\
 $c_5$ ($\mu$m$^{-1}$)	& $5.78$(f) &  $5.78\pm1.03$ & $6.5$(f) & $6.50\pm0.27$ \\
 $\gamma$ ($\mu$m$^{-1}$)& $0.82\pm0.27$ & $0.62\pm0.07$ & $1.23\pm1.04$  &  $0.91\pm0.12$ \\
 $R_V$ 	& $3.24\pm1.05$ & $3.19^{+0.86}_{-0.89}$  & $3.1$(f) & $3.1$(f) \\
 $x_0$ ($\mu$m$^{-1}$)	& $4.65\pm0.09$ &  $4.53\pm0.01$ & $4.65\pm0.05$ & $4.59\pm0.03$ \\
 $\beta$ & $0.60\pm0.03$ & $0.60\pm0.02$ & $0.80\pm0.12$ & $0.67\pm0.03$\\
  $A_V$  (mag)	& $0.52^{+0.13}_{-0.16}$ & $0.50^{+0.13}_{-0.10}$ & $2.12^{+0.68}_{-0.63}$ & $1.54^{+0.21}_{-0.22}$ \\ 
$\chi^2$/dof & 31/18 & 733/840 & 22/11 & 1268/1317 \\
\hline
\end{tabular}
\end{minipage}
\end{table}

\subsection{X-ray data}
For the two GRB afterglows the \emph{Swift} XRT data were downloaded from the \emph{Swift} data archive. The X-ray data were reduced using HEAsoft (version 6.10). GRB afterglow spectra were extracted in the 0.3--10.0 keV energy range using the latest calibration files. X-ray spectra were obtained near the time of the photometric data. The afterglow lightcurves were retrieved from the GRB light curve repository at the UK \emph{Swift} Science Data Centre, created as described in \citet{evans10}. The lightcurves were fitted by assuming a smoothly broken power-law \citep{beuermann}. Using the lightcurve fit, the X-ray spectra were then scaled to the relevant SED time by considering the photon weighted mean time of the X-ray spectra. We used the fitting procedure described in \citet{zafar} where the X-ray spectra are fitted and corrected for soft X-ray absorption below $\sim$ 3 keV within \texttt{XSPEC}. The spectrum of GRB\,080605 was fitted using a single power-law (PL) with a best fit photon index of $\Gamma=1.61\pm0.19$ and frozen for Galactic X-ray absorption of $6.67\times10^{20}$ cm$^{-2}$ (using the nH FTOOL; \citealt{kalberla}). The equivalent neutral hydrogen column density from the host galaxy of GRB\,080605, estimated from soft X-ray absorption is $N_{H,X}=5.58^{+0.44}_{-0.35}\times10^{21}$ cm$^{-2}$. The X-ray spectrum of GRB\,080805 was fitted with a best fit photon index of $\Gamma=1.82^{+0.25}_{-0.22}$ with fixed Galactic absorption ($3.46\times10^{20}$ cm$^{-2}$). The derived equivalent hydrogen column density is $N_{H,X}=1.22^{+0.35}_{-0.45}\times10^{22}$ cm$^{-2}$.

 \begin{figure}
	\centering 
	{\includegraphics[width=\columnwidth,clip=]{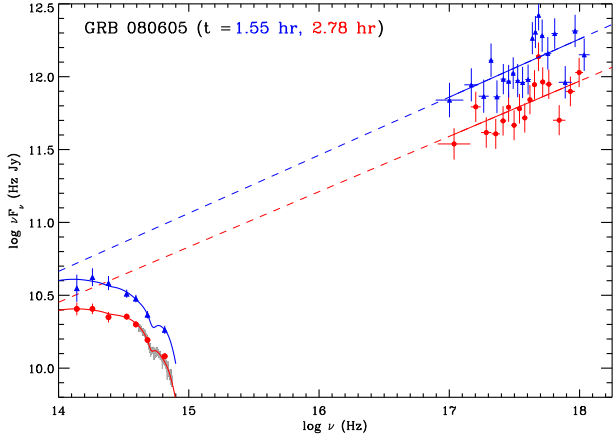} }
	\caption{NIR to X-ray SED of the afterglow of GRB\,080605 at 1.55 (blue triangles) and 2.78 hrs (red circles) after the burst. The grey curve represents the optical spectrum of the afterglow scaled to $t_0+2.78$ hrs. The solid lines represent the best-fits to the data. The dashed lines indicate the unextinguished power-laws.} 
		\label{sed080605} 
\end{figure}

\section{Results}\label{results}
We use a PL or broken power-law (BPL) to fit the SEDs and model the extinction with the  parameterized extinction curve defined in \citet{fm2} constitute of a UV linear component and a Drude component describing the UV/optical extinction curve in the rest frame of the object. In addition we used the $c_5$ parameter from \citet{fm3}. We will refer to the extinction model as FM. The parameterized extinction curve for $x>3.7$ $\mu$m$^{-1}$ is written as 
\begin{equation}
A_\lambda = A_V \left(\frac{1}{R_V}k(\lambda-V) + 1\right)
\end{equation}
where 
\[k(\lambda-V) = \left\{ 
\begin{array}{l l}
  c_1+c_2x+c_3 D(x,x_0,\gamma) & \quad \mbox{ $x\leq c_5$ }\\
  c_1+c_2x+c_3 D(x,x_0,\gamma)+c_4(x-c_5)^2 & \quad \mbox{ $x >c_5$ }\\ \end{array} \right. \]
where $x=\lambda^{-1}$ ($\mu$m$^{-1}$), and the Lorentzian-like Drude profile is expressed as
 \begin{eqnarray}
 D(x,x_0,\gamma) = \frac{x^2}{(x^2-x_0^2)^2+x^2\gamma^2} 
 \end{eqnarray}
where $x_0$ is the peak position, $\gamma$ is the bump width and $c_3$ is the bump strength. The UV linear component is controlled by the intercept $c_1$ and slope $c_2$. The extinction properties in the optical and infrared ($x<3.7$ $\mu$m$^{-1}$) are derived using spline interpolation. Additional useful quantities can be defined using the UV parameters e.g., $A_{\rm bump}=\pi c_3/(2\gamma)$ measures the area of the bump and $E_{\rm bump}=c_3/\gamma^2$ measures the maximum height of the bump above the UV linear extinction \citep{fm3}. SEDs of GRB afterglows are modelled with the parameterized extinction curve in the rest-frame of the GRB. 

We used the uncertainties on the data-points to create 1000 Monte Carlo (MC)
Gaussian random realizations. We fit these realizations, and in Table
\ref{table:1} we list the standard deviation of this distribution as
statistical errors on the best fit parameters. It should be noted as a
caveat, that the fitting parameters of the FM model are strongly correlated
\citep[see][for a detailed discussion]{fm3}, meaning that the
introduction of new data which may change the overall slope for example,
also impact the values of many of the other parameters, as observed in the
difference between the best-fit parameters here where we have new NIR data,
and \citet{zafar} for GRB\,080605 without this data. This caution is
applicable in this paper especially to GRB\,080805 where we have had to
assume a value for $R_V$.

\begin{figure}
	\centering 
	{\includegraphics[width=\columnwidth,clip=]{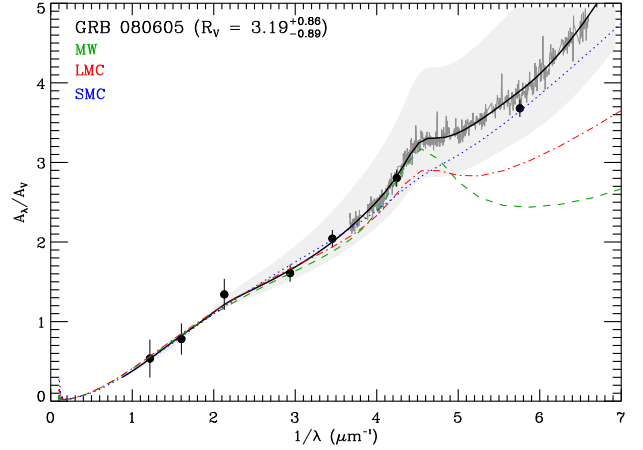} }
	\caption{Extinction curve of GRB\,080605 using the best-fit model at 2.78 hrs after the burst is shown in black line. The extinction curve is based on the best fit model given in Table \ref{table:1}. The grey curve represents the optical spectrum. The black circles correspond to the photometric data. The MW (green dashed curve), LMC (red dot-dashed curve) and SMC (blue dotted curve) models from \citet{pei} are also shown. The grey shaded area represents the $1\sigma$ uncertainty on the $R_V$ parameter.} 
		\label{ext080605} 
\end{figure} 

\subsection{SED of GRB\,080605}
The SED of the afterglow of GRB\,080605 is extracted at two epochs i.e.\ 1.55 and 2.78\,hrs after the burst (see Fig. \ref{sed080605}). The GROND-XRT and FORS2/GROND-XRT data were fitted at 1.55 and 2.78 hrs respectively. The FM parameterization is used to fit the afterglow SEDs (see \S \ref{results}). The data from both epochs were well fit with a single PL and a 2175\,\AA\ bump. The results of the fits are reported in Table \ref{table:1}. The model cannot constrain the $c_4$ and $c_5$ parameters at 1.55 hrs due to a lack of far-UV data, therefore, the values are fixed to the best fit value of 2.78 hrs. The best-fit parameters from both epochs are all consistent within the 90\% interval (see Table~\ref{table:1}). The extinction curve from the best-fit model at 2.78\,hrs is better-constrained, and so we use this to examine the extinction properties of the afterglow of GRB\,080605 (Fig. \ref{ext080605}). The extinction curve (in units of $A_\lambda$/$A_V$) of the GRB rises steeply into the UV like the SMC extinction curve but has a significant 2175\,\AA\ bump (see \S \ref{discussion}). The $A_{\rm bump}$  at 1.55 hrs epoch is $\sim$2$\sigma$ significant. We also fit the data with \citet{pei} SMC model and found that model without the bump is not a significantly better fit with an F-test probability of 90\% (SMC:$\chi^2$/dof=53/24, FM:$\chi^2$/dof=31/18). At 2.78 hrs epoch $A_{\rm bump}$ is $\sim 4$$\sigma$ significant. The derived metals-to-dust ratio based on 2.78 hrs epoch results is $N_{H,X}/A_V=1.12^{+0.16}_{-0.11}\times10^{22}$ cm$^{-2}$ mag$^{-1}$.

The FORS2-XRT SED of the afterglow was previously published in \citet{zafar} at 1.74 hrs after the burst, finding a large amount of extinction with $A_V=1.2\pm0.1$. Because of the lack of NIR data, the extinction curve was not constrained well in
that fit, resulting in a larger $A_V$ and relatively flatter extinction curve \citet{zafar}.
\citet{greiner10} implemented a GROND-XRT joint fit and found $A_V=0.47\pm0.03$ with the \citet{pei} MW dust extinction curve, similar to the value of $A_V$ found here.

\subsection{SED of GRB\,080805}
\begin{figure}
	\centering 
	{\includegraphics[width=\columnwidth,clip=]{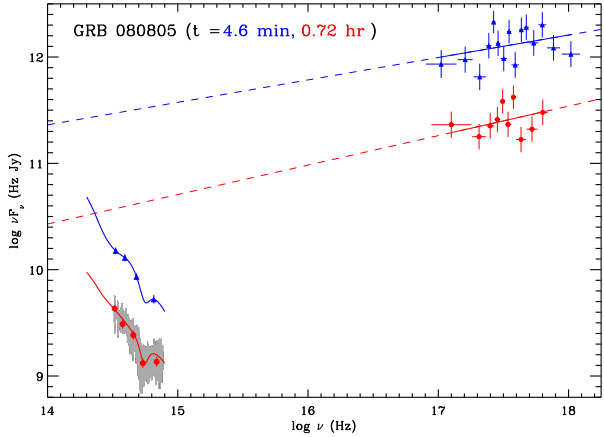} }
	\caption{Afterglow SED of GRB\,080805 at 4.6 min (blue triangles) and 0.72 hr (red circles). The grey curve corresponds to the FORS2 optical spectrum. The solid lines are the best-fits to the data. The dashed lines represent the unextinguished power-laws.} 
		\label{sed080805} 
\end{figure}

The SED of the afterglow of GRB\,080805 was constructed at 4.6 min and 0.72 hr after the burst. The GROND-XRT and FORS2-XRT SEDs are well fitted with a single PL and a 2175\,\AA\ bump (see Fig. \ref{sed080805}). The $R_V$ cannot be constrained for this burst because of the lack of NIR data, therefore, the value is fixed to the average value of the MW i.e. $R_V=3.1$. Because we do not have a detection in the far-UV at 4.6 min after the burst, the $c_4$ and $c_5$ fit parameters are fixed to the best-fit value of 0.72 hr epoch results. The $A_{\rm bump}$  at 0.72 hrs epoch is $\sim$5$\sigma$ significant. At 4.6 min epoch $A_{\rm bump}$  is $\approx$1$\sigma$ significant. Because of less available data in the optical at 4.6 min, the significance of the bump cannot confirmed at this epoch. We, therefore, rely mostly on 0.72 hrs epoch results for GRB\,080805. At 4.6 min after the burst \citet{greiner10} found that GROND-XRT data fit well with a BPL and \citet{pei} LMC dust model. We determine from the optical spectrum that the bump seen for GRB\,080805 is not LMC-like and has smaller strength and width to that of the \citet{pei} LMC bump. We fit the data with the FM extinction model using the 0.72 hr SED best-fit results as an initial guess and found that the SED can be well reproduced by a  single PL and required a 2175\,\AA\ bump. We also fit the data with \citet{pei} LMC model and found that a broken power-law is not a significantly better fit with an F-test probability of 83\% (PL:$\chi^2$/dof=25/16, BPL:$\chi^2$/dof=20/14).

\begin{figure}
	\centering 
	{\includegraphics[width=\columnwidth,clip=]{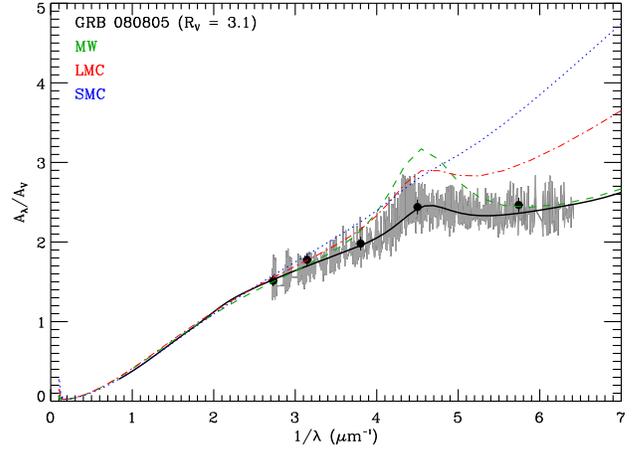} }
	\caption{Extinction curve of GRB\,080805 using the best-fit model at 0.72 hr after the burst (see  Table \ref{table:1}). The optical spectrum is illustrated by the grey curve. The black circles correspond to the photometric observations. The MW (green dashed curve), LMC (red dot-dashed curve) and SMC (blue dotted curve) models from \citet{pei} are also shown.} 
		\label{ext080805} 
\end{figure}

The extinction curve of the afterglow of GRB\,080805 was generated by using the best fit model obtained at 0.72 hr after the burst (see Fig. \ref{ext080805}). Due to the lack of the rest-frame optical/NIR data the overall slope of the extinction curve is not robust for this burst and can deviate from the one shown in Fig. \ref{ext080805}, resulting in a smaller or larger $A_V$. The inferred metals-to-dust ratio based on 0.72 hr epoch results is $N_{H,X}/A_V=0.79^{+0.24}_{-0.31}\times10^{22}$ cm$^{-2}$ mag$^{-1}$. It is also worth noting that \citet{schlegel} maps have been confirmed by \citet{dutra} up to $E(B-V)=0.25$\,mag. Assuming an uncertainty of 15\% \citep{schlafly}, we find that uncertainty in the Galactic extinction correction does not affect our results for both GRB afterglows and is always smaller than our statistical uncertainties.

\subsection{SED fitting of the GRB afterglow sample}
\begin{figure*}
	\centering 
	{\includegraphics[width=0.8\textwidth]{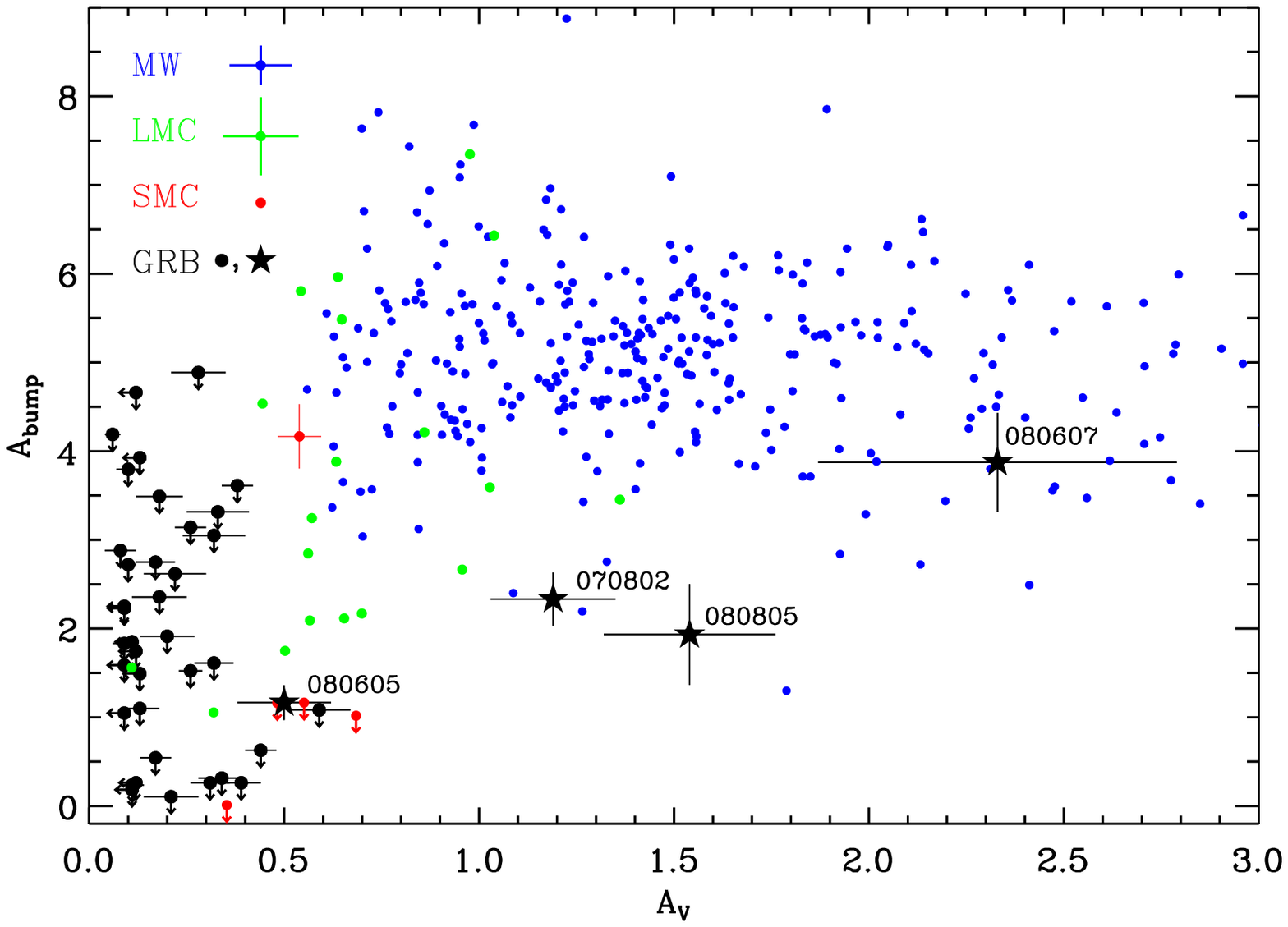} }
	\caption{$A_{\rm bump}$ versus $A_V$ for GRBs, the MW \citep{fm3}, LMC and SMC \citep{gordon}. The black circles indicate the $2\sigma$ $A_{\rm bump}$ upper limits for GRBs taken from the sample published in \citet{zafar}. The stars represents the GRB afterglows with a detection of the 2175\,\AA\ bump in their optical spectra. The blue, green and red points indicate the MW, LMC and SMC sightlines. The green and blue lines on the top-left corner correspond to the average errorbars for the LMC, and MW datapoints respectively.} 
		\label{strength} 
\end{figure*}

In this work, we re-fit the GRB afterglow data published in the spectroscopic sample study of \citet{zafar} with the FM extinction model. This was done to obtain the bump parameters especially $c_3$ (bump strength) and $\gamma$ (bump width) to study the 2175\,\AA\ bump properties i.e.: $i$) how common the 2175\,\AA\ bump is in GRB sightlines, $ii$) variation in the bump strength, width and area from one GRB sightline to another, and $iii$) do the GRB bump properties resemble those found in the Local Group? We fit SEDs of 38 GRB afterglows observed with the VLT/FORS instrument \citep[see][for the complete list of afterglows]{zafar}. All 38 GRBs in \citet{zafar} prefer an SMC-type extinction curve. We re-fit those GRBs with the FM extinction model and chose \citet{gordon} mean SMC parameters as an initial guess. From our FM fitting analysis we find that all GRBs with best fit SMC-type extinction curve in \citet{zafar} have insignificant values of $c_3$ ($<3\sigma$). We report $2\sigma$ upper limits for $c_3$ for all 38 afterglows (see Fig.~\ref{strength}). GRB\,070802 and GRB\,080607 are not re-fitted in the current analysis because of having been fit with the FM extinction model in \citet{zafar}. We use results for both afterglows from the analysis published in \citet{zafar}. GRB\,080605 and GRB\,080805 are, of course, reviewed in detail in this work and we use results from the current analysis (see Table~\ref{table:1}). 

%
%
\section{Discussion}\label{discussion}   

It has previously been found that GRBs reside in low-mass, faint,
sub-luminous, and blue galaxies \citep[e.g.,][]{lefloch03}. However recent
studies show that GRBs also occur in a population of dusty, luminous, red
and evolved galaxies \citep{piro,levan,berger,chen10,kruhler11}. This
suggests the previously known faint, young, and low-mass galaxy population is not representative of GRB
hosts as a whole. The paucity of the 2175\,\AA\ bump found in the
afterglows of GRBs to date seems likely to be an indication of our lack of
spectroscopic completeness due to dust bias \citep{zafar} and coincides
with the suggestion of \citet{noll07} that dust with a significant
2175\,\AA\ bump requires an evolved population.

The association between the carriers of the 2175\,\AA\ bump and evolved stellar populations would manifest itself also in the host galaxies of the respective GRBs. In comparison to the general population of GRB hosts, these galaxies should have higher stellar masses, higher IR luminosities and global metallicities. Such a trend is indeed observed in recent host samples, which indicate that, on average, galaxies hosting afterglows with a 2175\,\AA\ bump are redder, more massive and luminous than the standard GRB host \citep{kruhler11}. Similarly, the host of GRB\,080605 exhibits high gas-phase metallicity above 0.4 solar which puts it among the most metal rich GRB hosts ever detected \citep{thomas12}. Although number statistics and high-quality host observations are still sparse, the properties of the host galaxies hence seem to support the assertion that the presence of the 2175\,\AA\ bump traces environments with evolved stellar populations and substantial chemical enrichment.

Below we compare the results of this analysis to Local Group sightlines to
investigate the general properties of the dust in GRB hosts. We discuss the
extinction curve of GRB\,080605 which is surprisingly different from typical Local Group
extinction curves. \citet{cardelli} showed that Galactic sightlines could typically be 
well-fit with a $R_V$-dependent extinction curve. There are a few Galactic 
sightlines which are not adequately represented by the \citet{cardelli} 
extinction curve \citep{sofia}. We also investigate whether the bump is correlated 
with the presence of neutral carbon in the gas phase.


\subsection{Comparison with Local Group sightlines}

Using the results of the FM fitting analysis, we calculated the area and
maximum height of the bump for the GRB afterglow sample of \cite{zafar} by using
the relations described above (see \S~\ref{results}). In
Fig.~\ref{strength} we plot $A_{\rm bump}$ against $A_V$ and compare the
GRB afterglow sample results to the lines of sight in the MW \citep{fm3},
LMC and SMC \citep{gordon}. We find that $A_{\rm bump}$ (hence the strength
and the maximum height of the 2175\,\AA\ bump) for the four GRB afterglows
with spectroscopically detected 2175\,\AA\ bumps, is typically smaller than
for the vast majority of known Local Group sightlines for a given value of
$A_V$. In other words, the bump for GRB afterglows is not as prominent as
in lines of sight in the Local Group. It is interesting to note that it is
not simply that the bumps seen in GRB afterglows are weaker and shallower
than the bumps seen in the lines of sight in the MW---the extinction curves
of GRB afterglows are expected to be different from the MW extinction
curves---the bumps are also weaker than those detected in the Magellanic
Clouds, where one might expect the dust to be similar, because we are
probing actively star-forming environments rather than quiescent regions
\citep[see][]{gordon}.

The results of this work imply that the common usage of the average MW, LMC
and SMC extinction models to fit the GRB afterglow SEDs is inadequate \citep[see also][]{clayton00,gordon}. Here
we have shown that GRB sightlines with bumps can have weaker bumps than
either LMC or MW and bumps in combination with a steep extinction curve,
both highly uncharacteristic of most Local Group sightlines. In reality
Local Group sightlines exhibit a variety of extinction curves. This work
shows that, similar to the Local Group, GRB hosts seem to have a continuum
of dust extinction curves varying from steep to flat and bumpy to
featureless. In future, better rest-frame UV through NIR data will allow us
to obtain reasonable numbers of extinction curves of individual events
\citep[see \S~5.2 of][]{zafar}. In support of this objective the X-shooter
spectrograph on the VLT is now regularly obtaining UV through NIR spectra of
GRB afterglows \citep[e.g.,][]{antonio,delia}.

\begin{figure}
	\centering 
	{\includegraphics[width=\columnwidth,clip=]{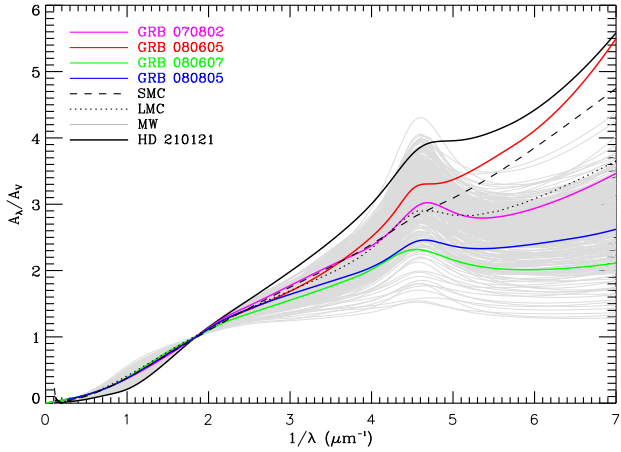} }
	\caption{Extinction curves of GRB\,070802 (magenta line),
                 GRB\,080605 (red line), GRB\,080607 (green line) and
                 GRB\,080805 (blue line). The SMC and LMC extinction curves
                 taken from \citet{pei} are shown in black dashed and dotted
                 lines respectively. The black solid line represents the
                 extinction towards the MW star HD\,210121. The extinction
                 curves of the MW sightlines taken from \citep{fm3} are
                 shown as grey lines.}
		\label{allext} 
\end{figure} 

\subsection{Dust composition}

The extinction curve of GRB\,080605 has a steeper rise into the UV and a
weak 2175\,\AA\ bump. Such extinction is also seen in the MW in the lines
of sight toward HD\,210121, and to some extent, HD\,62542, and HD\,204827,
all of which deviate substantially from the other MW extinction curves
\citep{valencic}. The two other interesting extinction curves in the MW are
lines of sight towards HD\,283809 and HD\,29647 showing a weaker bump but a
regular flat UV curvature \citep{clayton03,whittet}. Among these, the
sightline towards HD\,210121 is very unusual, displaying a very steep rise
into the far-UV, a small value of $R_V=2.4$ and a weak 2175\,\AA\ bump
(\citealt{larson,sofia}; see Fig.~\ref{allext}). \citet{weingartner01}
showed that their carbonaceous/silicate dust model can reproduce the
extinction towards HD\,210121. They suggested two ways of reproducing the
2175\,\AA\ bump: by introducing either small graphite particles or a
population of small carbonaceous grains including PAH molecules, and, more
likely, the rise into the far-UV may be due to a population of silicate grains
\citep[see][]{weingartner01}.

The relative weakness of the GRB afterglow bumps compared to Local Group sightlines hints
at the possibility that sightlines with a 2175\,\AA\ feature may be composed
of two distinct, physically separated dust populations, one with a steep
extinction curve and no bump, the other with a relatively flat extinction
curve and a strong bump. The steep, featureless curve would then dilute the
strength of the bump, and potentially produce the type of extinction curve
observed in GRB\,080605. In this scenario, both dust populations may reside
in the GRB host (e.g.\ in the molecular cloud in which the GRB forms and in
the more general host ISM), or one of the populations in a foreground
system. In the case of GRB\,070802, for example, we know that there is a
foreground absorber at lower redshift with a relatively high metal column
density \citep{ardis}, however, strong dust absorption in foreground systems
will not be the typical case. Rather, we have some evidence that two
distinct column densities exist in the gas phase within the host galaxy: one
nearby and one relatively distant
\citep{perna02,watson07,schady11,vreeswijk07,campana2009,kruhler11}. 
The unusual dust properties observed here may simply be a reflection of
this dual population.
\begin{figure}
	\centering 
	{\includegraphics[width=\columnwidth,clip=]{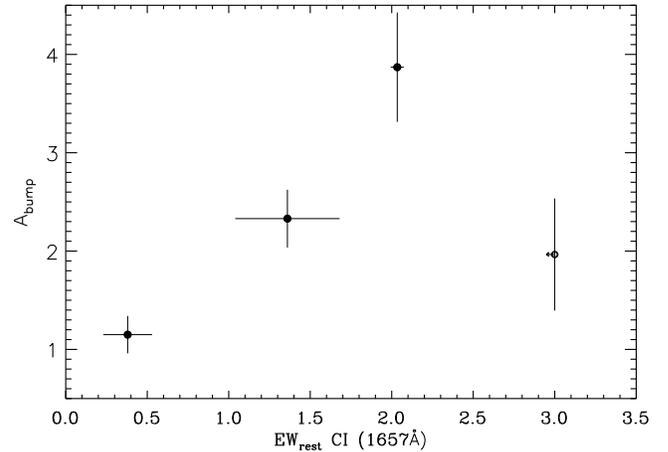} }
	\caption{$A_{\rm bump}$ against equivalent width of {C}\,{\sc i} $\lambda$1656.9 for GRB afterglows with the 2175\,\AA\ bump detected in their optical spectra. $2\sigma$ upper limit for the equivalent width of {C}\,{\sc i} is reported for GRB\,080805.} 
		\label{ciplot} 
\end{figure}

\subsection{The 2175\,\AA\ bump and {C}\,{\sc i}}

The first ionization potential of carbon is 11.3\,eV. It is not shielded by
neutral hydrogen and hence only expected to be present in regions with a low
density of ionizing photons. Carbonaceous materials are believed to be
responsible for the 2175\,\AA\ bump, and carbonaceous grain growth and
formation requires neutral carbon and molecules \citep{henning}. Therefore
it would not be surprising to see both features together from the same
environment. In the literature we find two detections of the {C}\,{\sc i}
absorption feature in the lines of sight towards the Local Group: $i)$ the line
of sight towards HD\,185418 in the MW \citep{sonnentrucker} and $ii)$ the line
of sight towards the SMC-bar with {C}\,{\sc i} detected in the MW but only
as an upper limit in the SMC \citep{welty}. Beyond the Local Group,
{C}\,{\sc i} absorption is detected in many systems \citep[e.g.,][]{ledoux02,jorgenson10}. \citet{junkkarinen} 
reported the detection of 2175\,\AA\ bump for an intervening damped Ly$\alpha$ absorber ($z= 0.524$) towards 
AO 0235+164 where {C}\,{\sc i} absorption line is also seen (see also \S 4.3 \citealt{ardis} for more discussion).

To our knowledge {C}\,{\sc i} is detected for all GRB afterglows with a
prominent 2175\,\AA\ bump except for GRB\,080805 due to its low
signal-to-noise ratio and redshift. Conversely, {C}\,{\sc i} is also
detected in the afterglow spectra of GRB\,060210 and GRB\,061121
\citep[see][]{fynbo}. However, GRB\,060210 is at $z=3.9133$, so that the
bump region is not covered in the optical spectrum. GRB\,061121 is at $z =
1.3145$ and has no detection of the 2175\,\AA\ bump. The spectra of both
afterglows appear to be blue suggesting little dust extinction for these
bursts. We also looked into the spectrum of GRB\,060418 which has a
reported intervening absorber at $z_{\rm abs}=1.107$ with a 2175\,\AA\ bump. At this redshift {C}\,{\sc i} is blended with the {C}\,{\sc iv} doublet
arising from the GRB host galaxy ($z=1.49$). Fitting the {C}\,{\sc iv}
doublet leaves a residual, suggesting the presence of another line, most
probably {C}{\sc i} from the intervening absorber.

It is striking that the extinguished afterglows with detected {C}\,{\sc i}
absorption also have a 2175\,\AA\ bump. The presence of {C}\,{\sc i}
absorption in the afterglows with the 2175\,\AA\ bump suggests that the UV
radiation field is weaker in these GRBs. In Fig.~\ref{ciplot} we attempt to
quantify this with the very little data we have. It is possible that there
may be a correlation between the area under the bump and the rest-frame
equivalent width of {C}\,{\sc i} $\lambda$1656.9. The numbers are too small,
however, to confirm any relation. Such a correlation would suggest that we
expect to see large equivalent widths of {C}\,{\sc i} for strong bumps and
less for weak ones. It should be noted that the ground electronic state of 
{C}\,{\sc i} split into three fine structure states as {C}\,{\sc i}, {C}\,{\sc i*}, and {C}\,{\sc i**}
 \citep[see e.g.,][]{jorgenson10}.
In our low resolution data we cannot distinguish the strength of the various contributions
of the excited states. If {C}\,{\sc i} line is not saturated then the EW of the complex is 
a useful indication of the column density of neutral carbon. But if {C}\,{\sc i} line is 
saturated and in the presence of significant excited states of {C}\,{\sc i}, our EW measurements 
may not be a meaningful measure of the column density in {C}\,{\sc i}.

We briefly also compared the area of the bump with the
underlying UV slope of the GRB extinction curves i.e.\ $c_2/R_V$. We find a
hint of a relationship between the two quantities suggesting that smaller
{C}\,{\sc i} equivalent widths are observed for GRBs with steeper UV
extinction curve slopes and vice versa. In future such relationships can be
checked with a larger sample of detected 2175\,\AA\ bump and significant
{C}\,{\sc i} absorption. If such a correlation holds then {C}\,{\sc i} could be used
as a spectroscopic signature to locate the 2175\,\AA\ bump in dusty
environments.

%
%
\section{Conclusions\label{conclusions}}

In this paper, we investigated the 2175\,\AA\ bump observed in GRB
afterglows. We performed multi-epoch NIR--X-ray SED analysis of GRB\,080605
and 080805 displaying 2175\,\AA\ bumps in their optical spectra. We find
the SEDs to be well fitted with a PL and an FM extinction model at different
epochs. So far the bump is spectroscopically detected in the spectra of five GRB afterglows
where one is in an intervening absorber. We compare the bump properties of
our GRB afterglow sample to Local Group sightlines. We find that $A_{\rm bump}$
for GRB afterglows is smaller for a given $A_V$ than almost all Local Group
sightlines. All four GRB extinction curves with detected 2175\,\AA\ bumps
differ from one another. The differences in the extinction curves suggest
that the use of the average MW, LMC and SMC extinction curve is inadequate. 
In particular, we know that the extinction curve of the afterglow of
GRB\,080605 is very different from the other GRB extinction curves, with a
2175\,\AA\ bump and steep rise into the far-UV. Such extinction curves and
the relative faintness of the bump strength with total extinction might
suggest that the dust we observe in the afterglow spectrum consists of two
different grain populations having different compositions. The presence of
the bump also contradicts the common notion that GRBs occur only in blue,
low-mass and faint galaxies. We find a hint of a possible relationship
between $A_{\rm bump}$ and neutral carbon for GRB afterglows that needs to
be further investigated.

\begin{acknowledgements} 
The Dark Cosmology Centre is funded by the Danish National Research Foundation. We are grateful to the anonymous referee for constructive comments. We are thankful to Daniele Malesani, Anja C. Andersen and Jens Hjorth for helpful discussion. JPUF acknowledges support form the ERC-StG grant EGGS-278202. GL is supported by the Carlsberg foundation. TK acknowledges support by the European Commission under the Marie Curie Intra-European Fellowship Programme and support by the DFG cluster of excellence `Origin and Structure of the Universe'. This work is based in part on observations done with the European Southern Observatory (ESO) utilizing the 8.2m VLT, Chile, under program 081.A-0856(B).
\end{acknowledgements}

\bibliography{bump.bib}

\begin{thebibliography}{68}
\expandafter\ifx\csname natexlab\endcsname\relax\def\natexlab#1{#1}\fi

\bibitem[{{Alard}(2000)}]{alard}
{Alard}, C. 2000, \aaps, 144, 363

\bibitem[{{Barthelmy} {et~al.}(2005){Barthelmy}, {Barbier}, {Cummings},
  {Fenimore}, {Gehrels}, {Hullinger}, {Krimm}, {Markwardt}, {Palmer},
  {Parsons}, {Sato}, {Suzuki}, {Takahashi}, {Tashiro}, \&
  {Tueller}}]{barthelmy}
{Barthelmy}, S.~D., {et~al.} 2005, \ssr, 120, 143

\bibitem[{{Berger} {et~al.}(2007){Berger}, {Fox}, {Kulkarni}, {Frail}, \&
  {Djorgovski}}]{berger}
{Berger}, E., {Fox}, D.~B., {Kulkarni}, S.~R., {Frail}, D.~A., \& {Djorgovski},
  S.~G. 2007, \apj, 660, 504

\bibitem[{{Beuermann} {et~al.}(1999){Beuermann}, {Hessman}, {Reinsch},
  {Nicklas}, {Vreeswijk}, {Galama}, {Rol}, {van Paradijs}, {Kouveliotou},
  {Frontera}, {Masetti}, {Palazzi}, \& {Pian}}]{beuermann}
{Beuermann}, K., {et~al.} 1999, \aap, 352, L26

\bibitem[{{Bloom} {et~al.}(2006){Bloom}, {Starr}, {Blake}, {Skrutskie}, \&
  {Falco}}]{bloom06}
{Bloom}, J.~S., {Starr}, D.~L., {Blake}, C.~H., {Skrutskie}, M.~F., \& {Falco},
  E.~E. 2006, in Astronomical Society of the Pacific Conference Series, Vol.
  351, Astronomical Data Analysis Software and Systems XV, ed. {C.~Gabriel,
  C.~Arviset, D.~Ponz, \& S.~Enrique}, 751

\bibitem[{{Buat} {et~al.}(2011){Buat}, {Giovannoli}, {Heinis}, {Charmandaris},
  {Coia}, {Daddi}, {Dickinson}, {Elbaz}, {Hwang}, {Morrison}, {Dasyra},
  {Aussel}, {Altieri}, {Dannerbauer}, {Kartaltepe}, {Leiton}, {Magdis},
  {Magnelli}, \& {Popesso}}]{buat}
{Buat}, V., {et~al.} 2011, \aap, 533, A93

\bibitem[{{Burrows} {et~al.}(2005){Burrows}, {Hill}, {Nousek}, {Kennea},
  {Wells}, {Osborne}, {Abbey}, {Beardmore}, {Mukerjee}, {Short}, {Chincarini},
  {Campana}, {Citterio}, {Moretti}, {Pagani}, {Tagliaferri}, {Giommi},
  {Capalbi}, {Tamburelli}, {Angelini}, {Cusumano}, {Br{\"a}uninger}, {Burkert},
  \& {Hartner}}]{burrows}
{Burrows}, D.~N., {et~al.} 2005, \ssr, 120, 165

\bibitem[{{Calzetti} {et~al.}(1994){Calzetti}, {Kinney}, \&
  {Storchi-Bergmann}}]{calzetti}
{Calzetti}, D., {Kinney}, A.~L., \& {Storchi-Bergmann}, T. 1994, \apj, 429, 582

\bibitem[{{Campana}(2009)}]{campana2009}
{Campana}, S. 2009, \apj, 699, 1144

\bibitem[{{Cardelli} \& {Clayton}(1991)}]{cardelli}
{Cardelli}, J.~A., \& {Clayton}, G.~C. 1991, \aj, 101, 1021

\bibitem[{{Chen} {et~al.}(2010){Chen}, {Perley}, {Wilson}, {Cenko}, {Levan},
  {Bloom}, {Prochaska}, {Tanvir}, {Dessauges-Zavadsky}, \& {Pettini}}]{chen10}
{Chen}, H., {et~al.} 2010, \apjl, 723, L218

\bibitem[{{Christensen} {et~al.}(2011){Christensen}, {Fynbo}, {Prochaska},
  {Th{\"o}ne}, {de Ugarte Postigo}, \& {Jakobsson}}]{christensen11}
{Christensen}, L., {Fynbo}, J.~P.~U., {Prochaska}, J.~X., {Th{\"o}ne}, C.~C.,
  {de Ugarte Postigo}, A., \& {Jakobsson}, P. 2011, \apj, 727, 73

\bibitem[{{Clayton} {et~al.}(2003){Clayton}, {Gordon}, {Salama}, {Allamandola},
  {Martin}, {Snow}, {Whittet}, {Witt}, \& {Wolff}}]{clayton03}
{Clayton}, G.~C., {et~al.} 2003, \apj, 592, 947

\bibitem[{{Clayton} {et~al.}(2000){Clayton}, {Gordon}, \& {Wolff}}]{clayton00}
{Clayton}, G.~C., {Gordon}, K.~D., \& {Wolff}, M.~J. 2000, \apjs, 129, 147

\bibitem[{{de Ugarte Postigo} {et~al.}(2010){de Ugarte Postigo}, {Goldoni},
  {Th{\"o}ne}, {Vergani}, {D'Elia}, {Piranomonte}, {Malesani}, {Covino},
  {Flores}, {Fynbo}, {Hjorth}, {Wijers}, {D'Odorico}, {Hammer}, {Kaper},
  {Kj{\ae}rgaard}, {Randich}, {Andersen}, {Antonelli}, {Christensen},
  {D'Avanzo}, {Fiore}, {Groot}, {Maiorano}, {Palazzi}, {Pian}, {Tagliaferri},
  {van den Ancker}, \& {Vernet}}]{antonio}
{de Ugarte Postigo}, A., {et~al.} 2010, \aap, 513, A42

\bibitem[{{D'Elia} {et~al.}(2010){D'Elia}, {Fynbo}, {Covino}, {Goldoni},
  {Jakobsson}, {Matteucci}, {Piranomonte}, {Sollerman}, {Th{\"o}ne}, {Vergani},
  {Vreeswijk}, {Watson}, {Wiersema}, {Zafar}, {de Ugarte Postigo}, {Flores},
  {Hjorth}, {Kaper}, {Levan}, {Malesani}, {Milvang-Jensen}, {Pian},
  {Tagliaferri}, \& {Tanvir}}]{delia}
{D'Elia}, V., {et~al.} 2010, \aap, 523, A36

\bibitem[{{Draine}(2003)}]{draine03}
{Draine}, B.~T. 2003, \araa, 41, 241

\bibitem[{{Dutra} {et~al.}(2003){Dutra}, {Ahumada}, {Clari{\'a}}, {Bica}, \&
  {Barbuy}}]{dutra}
{Dutra}, C.~M., {Ahumada}, A.~V., {Clari{\'a}}, J.~J., {Bica}, E., \& {Barbuy},
  B. 2003, \aap, 408, 287

\bibitem[{{El{\'{\i}}asd{\'o}ttir} {et~al.}(2009){El{\'{\i}}asd{\'o}ttir},
  {Fynbo}, {Hjorth}, {Ledoux}, {Watson}, {Andersen}, {Malesani}, {Vreeswijk},
  {Prochaska}, {Sollerman}, \& {Jaunsen}}]{ardis}
{El{\'{\i}}asd{\'o}ttir}, {\'A}., {et~al.} 2009, \apj, 697, 1725

\bibitem[{{Ellison} {et~al.}(2006){Ellison}, {Vreeswijk}, {Ledoux}, {Willis},
  {Jaunsen}, {Wijers}, {Smette}, {Fynbo}, {M{\o}ller}, {Hjorth}, \&
  {Kaufer}}]{ellison}
{Ellison}, S.~L., {et~al.} 2006, \mnras, 372, L38

\bibitem[{{Evans} {et~al.}(2010){Evans}, {Willingale}, {Osborne}, {O'Brien},
  {Page}, {Markwardt}, {Barthelmy}, {Beardmore}, {Burrows}, {Pagani},
  {Starling}, {Gehrels}, \& {Romano}}]{evans10}
{Evans}, P.~A., {et~al.} 2010, \aap, 519, A102

\bibitem[{{Fitzpatrick} \& {Massa}(1990)}]{fm2}
{Fitzpatrick}, E.~L., \& {Massa}, D. 1990, \apjs, 72, 163

\bibitem[{{Fitzpatrick} \& {Massa}(2007)}]{fm3}
---. 2007, \apj, 663, 320

\bibitem[{{Fynbo} {et~al.}(2009){Fynbo}, {Jakobsson}, {Prochaska}, {Malesani},
  {Ledoux}, {de Ugarte Postigo}, {Nardini}, {Vreeswijk}, {Wiersema}, {Hjorth},
  {Sollerman}, {Chen}, {Th{\"o}ne}, {Bj{\"o}rnsson}, {Bloom}, {Castro-Tirado},
  {Christensen}, {De Cia}, {Fruchter}, {Gorosabel}, {Graham}, {Jaunsen},
  {Jensen}, {Kann}, {Kouveliotou}, {Levan}, {Maund}, {Masetti},
  {Milvang-Jensen}, {Palazzi}, {Perley}, {Pian}, {Rol}, {Schady}, {Starling},
  {Tanvir}, {Watson}, {Xu}, {Augusteijn}, {Grundahl}, {Telting}, \&
  {Quirion}}]{fynbo}
{Fynbo}, J.~P.~U., {et~al.} 2009, \apjs, 185, 526

\bibitem[{{Gordon} {et~al.}(2003){Gordon}, {Clayton}, {Misselt}, {Landolt}, \&
  {Wolff}}]{gordon}
{Gordon}, K.~D., {Clayton}, G.~C., {Misselt}, K.~A., {Landolt}, A.~U., \&
  {Wolff}, M.~J. 2003, \apj, 594, 279

\bibitem[{{Greiner} {et~al.}(2011){Greiner}, {Kr{\"u}hler}, {Klose}, {Afonso},
  {Clemens}, {Filgas}, {Hartmann}, {K{\"u}pc{\"u} Yolda{\c s}}, {Nardini},
  {Olivares E.}, {Rau}, {Rossi}, {Schady}, \& {Updike}}]{greiner10}
{Greiner}, J., {et~al.} 2011, \aap, 526, A30

\bibitem[{{Henning} \& {Salama}(1998)}]{henning}
{Henning}, T., \& {Salama}, F. 1998, Science, 282, 2204

\bibitem[{{Henrard} {et~al.}(1997){Henrard}, {Lambin}, \& {Lucas}}]{henard}
{Henrard}, L., {Lambin}, P., \& {Lucas}, A.~A. 1997, \apj, 487, 719

\bibitem[{{Jakobsson} {et~al.}(2008{\natexlab{a}}){Jakobsson}, {Fynbo},
  {Vreeswijk}, \& {de Ugarte Postigo}}]{jakobsson084}
{Jakobsson}, P., {Fynbo}, J.~P.~U., {Vreeswijk}, P.~M., \& {de Ugarte Postigo},
  A. 2008{\natexlab{a}}, GRB Coordinates Network, 8077, 1

\bibitem[{{Jakobsson} {et~al.}(2008{\natexlab{b}}){Jakobsson}, {Vreeswijk},
  {Xu}, \& {Thoene}}]{jakobsson083}
{Jakobsson}, P., {Vreeswijk}, P.~M., {Xu}, D., \& {Thoene}, C.~C.
  2008{\natexlab{b}}, GRB Coordinates Network, 7832, 1

\bibitem[{{Jiang} {et~al.}(2011){Jiang}, {Ge}, {Zhou}, {Wang}, \&
  {Wang}}]{jiang}
{Jiang}, P., {Ge}, J., {Zhou}, H., {Wang}, J., \& {Wang}, T. 2011, \apj, 732,
  110

\bibitem[{{Jorgenson} {et~al.}(2010){Jorgenson}, {Wolfe}, \&
  {Prochaska}}]{jorgenson10}
{Jorgenson}, R.~A., {Wolfe}, A.~M., \& {Prochaska}, J.~X. 2010, \apj, 722, 460

\bibitem[{{Junkkarinen} {et~al.}(2004){Junkkarinen}, {Cohen}, {Beaver},
  {Burbidge}, {Lyons}, \& {Madejski}}]{junkkarinen}
{Junkkarinen}, V.~T., {Cohen}, R.~D., {Beaver}, E.~A., {Burbidge}, E.~M.,
  {Lyons}, R.~W., \& {Madejski}, G. 2004, \apj, 614, 658

\bibitem[{{Kalberla} {et~al.}(2005){Kalberla}, {Burton}, {Hartmann}, {Arnal},
  {Bajaja}, {Morras}, \& {P{\"o}ppel}}]{kalberla}
{Kalberla}, P.~M.~W., {Burton}, W.~B., {Hartmann}, D., {Arnal}, E.~M.,
  {Bajaja}, E., {Morras}, R., \& {P{\"o}ppel}, W.~G.~L. 2005, \aap, 440, 775

\bibitem[{{Kr{\"u}hler} {et~al.}(2012){Kr{\"u}hler}, {Fynbo}, {Geier},
  {Hjorth}, {Malesani}, {Milvang-Jensen}, {Sparre}, {Watson}, \&
  {Zafar}}]{thomas12}
{Kr{\"u}hler}, T., {et~al.} 2012, \aap~submitted~arXiv:1203.1919

\bibitem[{{Kr{\"u}hler} {et~al.}(2011){Kr{\"u}hler}, {Greiner}, {Schady},
  {Savaglio}, {Afonso}, {Clemens}, {Elliott}, {Filgas}, {Gruber}, {Kann},
  {Klose}, {K{\"u}pc{\"u}-Yolda{\c s}}, {McBreen}, {Olivares}, {Pierini},
  {Rau}, {Rossi}, {Nardini}, {Nicuesa Guelbenzu}, {Sudilovsky}, \&
  {Updike}}]{kruhler11}
---. 2011, \aap, 534, A108

\bibitem[{{Kr{\"u}hler} {et~al.}(2008){Kr{\"u}hler}, {K{\"u}pc{\"u} Yolda{\c
  s}}, {Greiner}, {Clemens}, {McBreen}, {Primak}, {Savaglio}, {Yolda{\c s}},
  {Szokoly}, \& {Klose}}]{kruhler08}
---. 2008, \apj, 685, 376

\bibitem[{{Larson} {et~al.}(2000){Larson}, {Wolff}, {Roberge}, {Whittet}, \&
  {He}}]{larson}
{Larson}, K.~A., {Wolff}, M.~J., {Roberge}, W.~G., {Whittet}, D.~C.~B., \&
  {He}, L. 2000, \apj, 532, 1021

\bibitem[{{Le Floc'h} {et~al.}(2003){Le Floc'h}, {Duc}, {Mirabel}, {Sanders},
  {Bosch}, {Diaz}, {Donzelli}, {Rodrigues}, {Courvoisier}, {Greiner},
  {Mereghetti}, {Melnick}, {Maza}, \& {Minniti}}]{lefloch03}
{Le Floc'h}, E., {et~al.} 2003, \aap, 400, 499

\bibitem[{{Ledoux} {et~al.}(2002){Ledoux}, {Srianand}, \&
  {Petitjean}}]{ledoux02}
{Ledoux}, C., {Srianand}, R., \& {Petitjean}, P. 2002, \aap, 392, 781

\bibitem[{{Lequeux} {et~al.}(1982){Lequeux}, {Maurice}, {Prevot-Burnichon},
  {Prevot}, \& {Rocca-Volmerange}}]{lequeux}
{Lequeux}, J., {Maurice}, E., {Prevot-Burnichon}, M., {Prevot}, L., \&
  {Rocca-Volmerange}, B. 1982, \aap, 113, L15

\bibitem[{{Levan} {et~al.}(2006){Levan}, {Fruchter}, {Rhoads}, {Mobasher},
  {Tanvir}, {Gorosabel}, {Rol}, {Kouveliotou}, {Dell'Antonio}, {Merrill},
  {Bergeron}, {Castro Cer{\'o}n}, {Masetti}, {Vreeswijk}, {Antonelli},
  {Bersier}, {Castro-Tirado}, {Fynbo}, {Garnavich}, {Holland}, {Hjorth},
  {Nugent}, {Pian}, {Smette}, {Thomsen}, {Thorsett}, \& {Wijers}}]{levan}
{Levan}, A., {et~al.} 2006, \apj, 647, 471

\bibitem[{{Mathis}(1994)}]{mathis94}
{Mathis}, J.~S. 1994, \apj, 422, 176

\bibitem[{{Nandy} {et~al.}(1981){Nandy}, {Morgan}, {Willis}, {Wilson}, \&
  {Gondhalekar}}]{nandy}
{Nandy}, K., {Morgan}, D.~H., {Willis}, A.~J., {Wilson}, R., \& {Gondhalekar},
  P.~M. 1981, \mnras, 196, 955

\bibitem[{{Noll} {et~al.}(2007){Noll}, {Pierini}, {Pannella}, \&
  {Savaglio}}]{noll07}
{Noll}, S., {Pierini}, D., {Pannella}, M., \& {Savaglio}, S. 2007, \aap, 472,
  455

\bibitem[{{Noterdaeme} {et~al.}(2009){Noterdaeme}, {Ledoux}, {Srianand},
  {Petitjean}, \& {Lopez}}]{noterdaeme}
{Noterdaeme}, P., {Ledoux}, C., {Srianand}, R., {Petitjean}, P., \& {Lopez}, S.
  2009, \aap, 503, 765

\bibitem[{{Pei}(1992)}]{pei}
{Pei}, Y.~C. 1992, \apj, 395, 130

\bibitem[{{Perley} {et~al.}(2011){Perley}, {Morgan}, {Updike}, {Yuan},
  {Akerlof}, {Miller}, {Bloom}, {Cenko}, {Li}, {Filippenko}, {Prochaska},
  {Kann}, {Tanvir}, {Levan}, {Butler}, {Christian}, {Hartmann}, {Milne},
  {Rykoff}, {Rujopakarn}, {Wheeler}, \& {Williams}}]{perley09}
{Perley}, D.~A., {et~al.} 2011, \aj, 141, 36

\bibitem[{{Perna} \& {Lazzati}(2002)}]{perna02}
{Perna}, R., \& {Lazzati}, D. 2002, \apj, 580, 261

\bibitem[{{Piro} {et~al.}(2001){Piro}, {Garmire}, {Garcia}, {Antonelli},
  {Costa}, {Feroci}, {Frail}, {Harrison}, {Hurley}, {M{\'e}sz{\'a}ros}, \&
  {Waxman}}]{piro}
{Piro}, L., {et~al.} 2001, \apj, 558, 442

\bibitem[{{Prochaska} {et~al.}(2009){Prochaska}, {Sheffer}, {Perley}, {Bloom},
  {Lopez}, {Dessauges-Zavadsky}, {Chen}, {Filippenko}, {Ganeshalingam}, {Li},
  {Miller}, \& {Starr}}]{prochaska09}
{Prochaska}, J.~X., {et~al.} 2009, \apjl, 691, L27

\bibitem[{{Roming} {et~al.}(2005){Roming}, {Kennedy}, {Mason}, {Nousek}, {Ahr},
  {Bingham}, {Broos}, {Carter}, {Hancock}, {Huckle}, {Hunsberger}, {Kawakami},
  {Killough}, {Koch}, {McLelland}, {Smith}, {Smith}, {Soto}, {Boyd},
  {Breeveld}, {Holland}, {Ivanushkina}, {Pryzby}, {Still}, \&
  {Stock}}]{roming05}
{Roming}, P.~W.~A., {et~al.} 2005, \ssr, 120, 95

\bibitem[{{Schady} {et~al.}(2011){Schady}, {Savaglio}, {Kr{\"u}hler},
  {Greiner}, \& {Rau}}]{schady11}
{Schady}, P., {Savaglio}, S., {Kr{\"u}hler}, T., {Greiner}, J., \& {Rau}, A.
  2011, \aap, 525, A113

\bibitem[{{Schlafly} {et~al.}(2010){Schlafly}, {Finkbeiner}, {Schlegel},
  {Juri{\'c}}, {Ivezi{\'c}}, {Gibson}, {Knapp}, \& {Weaver}}]{schlafly}
{Schlafly}, E.~F., {Finkbeiner}, D.~P., {Schlegel}, D.~J., {Juri{\'c}}, M.,
  {Ivezi{\'c}}, {\v Z}., {Gibson}, R.~R., {Knapp}, G.~R., \& {Weaver}, B.~A.
  2010, \apj, 725, 1175

\bibitem[{{Schlegel} {et~al.}(1998){Schlegel}, {Finkbeiner}, \&
  {Davis}}]{schlegel}
{Schlegel}, D.~J., {Finkbeiner}, D.~P., \& {Davis}, M. 1998, \apj, 500, 525

\bibitem[{{Sofia} {et~al.}(2005){Sofia}, {Wolff}, {Rachford}, {Gordon},
  {Clayton}, {Cartledge}, {Martin}, {Draine}, {Mathis}, {Snow}, \&
  {Whittet}}]{sofia}
{Sofia}, U.~J., {et~al.} 2005, \apj, 625, 167

\bibitem[{{Sonnentrucker} {et~al.}(2003){Sonnentrucker}, {Friedman}, {Welty},
  {York}, \& {Snow}}]{sonnentrucker}
{Sonnentrucker}, P., {Friedman}, S.~D., {Welty}, D.~E., {York}, D.~G., \&
  {Snow}, T.~P. 2003, \apj, 596, 350

\bibitem[{{Stecher}(1965)}]{stecher65}
{Stecher}, T.~P. 1965, \apj, 142, 1683

\bibitem[{{Steel} \& {Duley}(1987)}]{steel}
{Steel}, T.~M., \& {Duley}, W.~W. 1987, \apj, 315, 337

\bibitem[{{Valencic} {et~al.}(2003){Valencic}, {Clayton}, {Gordon}, \&
  {Smith}}]{valencic}
{Valencic}, L.~A., {Clayton}, G.~C., {Gordon}, K.~D., \& {Smith}, T.~L. 2003,
  \apj, 598, 369

\bibitem[{{Vreeswijk} {et~al.}(2007){Vreeswijk}, {Ledoux}, {Smette}, {Ellison},
  {Jaunsen}, {Andersen}, {Fruchter}, {Fynbo}, {Hjorth}, {Kaufer}, {M{\o}ller},
  {Petitjean}, {Savaglio}, \& {Wijers}}]{vreeswijk07}
{Vreeswijk}, P.~M., {et~al.} 2007, \aap, 468, 83

\bibitem[{{Wang} {et~al.}(2004){Wang}, {Hall}, {Ge}, {Li}, \&
  {Schneider}}]{wang}
{Wang}, J., {Hall}, P.~B., {Ge}, J., {Li}, A., \& {Schneider}, D.~P. 2004,
  \apj, 609, 589

\bibitem[{{Watson} {et~al.}(2007){Watson}, {Hjorth}, {Fynbo}, {Jakobsson},
  {Foley}, {Sollerman}, \& {Wijers}}]{watson07}
{Watson}, D., {Hjorth}, J., {Fynbo}, J.~P.~U., {Jakobsson}, P., {Foley}, S.,
  {Sollerman}, J., \& {Wijers}, R.~A.~M.~J. 2007, \apjl, 660, L101

\bibitem[{{Weingartner} \& {Draine}(2001)}]{weingartner01}
{Weingartner}, J.~C., \& {Draine}, B.~T. 2001, \apj, 548, 296

\bibitem[{{Welty} {et~al.}(1997){Welty}, {Lauroesch}, {Blades}, {Hobbs}, \&
  {York}}]{welty}
{Welty}, D.~E., {Lauroesch}, J.~T., {Blades}, J.~C., {Hobbs}, L.~M., \& {York},
  D.~G. 1997, \apj, 489, 672

\bibitem[{{Whittet} {et~al.}(2004){Whittet}, {Shenoy}, {Clayton}, \&
  {Gordon}}]{whittet}
{Whittet}, D.~C.~B., {Shenoy}, S.~S., {Clayton}, G.~C., \& {Gordon}, K.~D.
  2004, \apj, 602, 291

\bibitem[{{Zafar} {et~al.}(2011){Zafar}, {Watson}, {Fynbo}, {Malesani},
  {Jakobsson}, \& {de Ugarte Postigo}}]{zafar}
{Zafar}, T., {Watson}, D., {Fynbo}, J.~P.~U., {Malesani}, D., {Jakobsson}, P.,
  \& {de Ugarte Postigo}, A. 2011, \aap, 532, A143

\bibitem[{{Zhou} {et~al.}(2010){Zhou}, {Ge}, {Lu}, {Wang}, {Yuan}, {Jiang}, \&
  {Shan}}]{zhou}
{Zhou}, H., {Ge}, J., {Lu}, H., {Wang}, T., {Yuan}, W., {Jiang}, P., \& {Shan},
  H. 2010, \apj, 708, 742

\end{thebibliography}

\end{document}